\documentclass[notitlepage,reprint,nofootinbib,preprintnumbers,twocolumn,floatfix,superscriptaddress,pra,showpacs]{revtex4-2}
\usepackage{physics}
\usepackage[clock]{ifsym}
\usepackage{natbib}
\usepackage{amsmath,amssymb}
\usepackage{graphicx}
\usepackage[dvipsnames]{xcolor}
\usepackage[super]{nth}
\usepackage[caption=false]{subfig}
\usepackage{float}
\usepackage[pdftex]{hyperref} 

\begin{document}

 \title{Testing quantum theory on curved space-time with quantum networks}
	\author{Johannes Borregaard}
    \email[]{borregaard@fas.harvard.edu}
	\affiliation{Department of Physics, Harvard University, Cambridge, Massachusetts 02138, USA}
	\author{Igor Pikovski}
    \email[]{pikovski@stevens.edu}
	\affiliation{Department of Physics, Stockholm University, SE-106 91 Stockholm, Sweden}
	\affiliation{Department of Physics, Stevens Institute of Technology, Hoboken, NJ 07030, USA}
 \begin{abstract}   
     Quantum technologies present new opportunities for fundamental tests of nature. One potential application is to probe the interplay between quantum physics and general relativity -- a field of physics with no empirical evidence yet. Here we show that quantum networks open a new window to test this interface. We demonstrate how photon mediated entanglement between atomic or atom-like systems can be used to probe time-dilation induced entanglement and interference modulation. Key are non-local measurements between clocks in a gravitational field, which can be achieved either through direct photon interference or by using auxiliary entanglement. The resulting observable depends on the interference between different proper times, and can only be explained if both quantum theory and general relativity are taken into account. The proposed protocol enables clock interferometry on km-scale separations and beyond. Our work thus shows a realistic experimental route for a first test of quantum theory on curved space-time, opening up new scientific opportunities for quantum networks.
 \end{abstract} 
 
 \maketitle

\section{Introduction}
The interface between quantum theory and gravity remains one of the main open questions in physics. General relativity has been confirmed in a vast range of observations \cite{will2018theory}. However, effects where both, quantum theory and general relativity have to simultaneously describe the same degrees of freedom, remain untested. So far, experiments have only accessed the Newtonian limit of gravity in the quantum domain, such as through phase-shift measurements in matter-wave interferometry \cite{colella1975observation, muller2010precision, overstreet2022observation, tino2021testing}, or quantum jumps of neutrons in a gravitational potential \cite{nesvizhevsky2002quantum}. However, going beyond the Newtonian limit is of fundamental interest: Gravity is the only known interaction that originates from curvature of space-time, and is thus fundamentally different from other forces. Experiments where quantum phenomena are altered due to the space-time curvature, and which cannot be described simply by a Newtonian potential, would provide a genuine test of general relativistic coupling to quantum matter. Such tests would amount to probing quantum theory on curved space-time, but so far they remain unrealized due to daunting experimental requirements. 

Here we show that entangled clocks can test quantum theory on curved space-time, and that such experiments are within reach of current technology. Our results demonstrate the strength of sensing with entangled systems~\cite{guo2020distributed,Zhang2021,Jun2024} and non-local quantum measurements, which enable a test of gravity that cannot be achieved with any classical sensor. Our work builds on \textit{proper time interferometry}~\cite{Zych2011} and extends it to quantum networks of entangled systems~\cite{komar2014quantum,pompili2021,Daiss2021,Knaut2024}. The key feature is that the time-evolution of a single clock system is governed by different proper times in superposition, which causes entanglement of spatial and clock degrees of freedom.  
Our work extends these considerations to entangled clocks and we show that interference of proper times across large distances and space-time curvatures can be achieved. Our approach combines non-local clock states with non-local readout, which can be achieved by means of distributing entanglement between local nodes as envisioned in a quantum network. Consequently, quantum networking techniques for entanglement distribution such as loss-tolerant, heralded entanglement generation~\cite{Beukers2024} and quantum repeaters~\cite{Azuma2023} can be used to increase the scale of the proper time interferometry. In addition, we discuss a specific implementation with emission-based spin-photon entanglement generation suitable for current state-of-the-art atomic and atomic-like systems~\cite{Madjarov2019,pompili2021,Daiss2021,nichol2022elementary,Knaut2024,liu2024,stolk2024}, thus showing how tests of quantum theory on curved space-time can be achieved with well-developed techniques from quantum optics and quantum communication.

	\section{Matter-wave interferometry}
Many remarkable experiments with atomic clocks have demonstrated the gravitational redshift \cite{hafele1972around}, now down to mm-scales on Earth~\cite{zheng2023lab,zheng2022differential,bothwell2022resolving}. In essence such experiments demonstrate the relevance of proper time even in low-energy quantum theory. To account for the correct clock rates, in their respective co-moving frames, the atomic clocks must evolve according to their local unitary $U_c(\tau)=e^{-i H_c \tau/\hbar}$, where $H_c$ is the local Hamiltonian responsible for the evolution of the atomic states, and $\tau$ is the proper time as experienced by the clocks' world-lines. Time dilation arises because different clocks experience different proper times due to the gravitational potential $\Phi=-GM/r$  and the momentum $p$ of the atoms: $\tau \approx \left(1+\frac{\Phi}{c^2} - \frac{p^2}{m^2 c^2} \right)t$. This is in stark contrast to Newtonian physics where a single universal coordinate time $t$ describes all systems. 

A natural extension to further probe quantum features interfacing gravity is thus to ask: Can evolution with respect to a \textit{superposition} of different proper times occur? Gedankenexperiments of this kind for special relativistic time dilation have, to our knowledge, first been proposed in Ref.~\cite{vedral2008schrodinger}. For gravitational time dilation, this opens a new route to probe the interplay of quantum theory and general relativity, first proposed in Ref. \cite{Zych2011} and further studied in a series of theoretical works~\cite{zych2012general, Pikovski2015, pikovski2017time, anastopoulos2018equivalence, zhou2018quantum, schwartz2019post, zych2019gravitational, paige2020classical, khandelwal2020universal,hartong2024coupling}. Such proper time interferometry can demonstrate post-Newtonian influence on genuine quantum phenomena on the one hand, and constrain speculative models of new physics such as the existence of intrinsic fundamental clocks \cite{Zych2011,sinha2011atom}. Phenomena that require both quantum theory and general relativity arise as internal states and the world-lines are both in a quantum superposition. This can lead to new quantum effects at the interface with relativity, such as time dilation induced decoherence \cite{Pikovski2015}, shifts in atomic frequency \cite{haustein2019mass, smith2020quantum}, quantum effects due to boosts \cite{paige2020classical}, or an interplay with many-body synchronization \cite{chu2024exploring}. However, apart from an experimental demonstration that simulated the effect \cite{margalit2015self}, such proper time interference phenomena in clocks still remain outside the reach of current experiments. Proposals for its implementation have mainly focused on variations of the matter-wave scheme \cite{Zych2011, bushev2016single, loriani2019interference, roura2020gravitational} but the limited spatial extent of achievable superpositions and the limited coherence times make the experimental implementation challenging. An extension of such tests to photons was also proposed \cite{zych2012general}, where superpositions of photon wave-packets would undergo different Shapiro-delays rather than different proper time evolutions. The challenge in such photonic interferometry tests, essentially photonic versions of the COW effect \cite{colella1975observation}, is the required distance or delay line for propagating photons, which has not been achieved yet but has been considered for implementation in different scenarios \cite{rideout2012fundamental,brodutch2015post,pallister2017blueprint,mazzarella2021goals,sidhu2021advances,lu2022micius,lohrmann2023classical,barzel2024entanglement, wu2024single}. Notably, a recent study has shown how quantum memories can replace photonic delay lines and thus assist photonic COW tests \cite{barzel2024entanglement}.  

\begin{figure*} [t]
\centering
\includegraphics[width=1\textwidth]{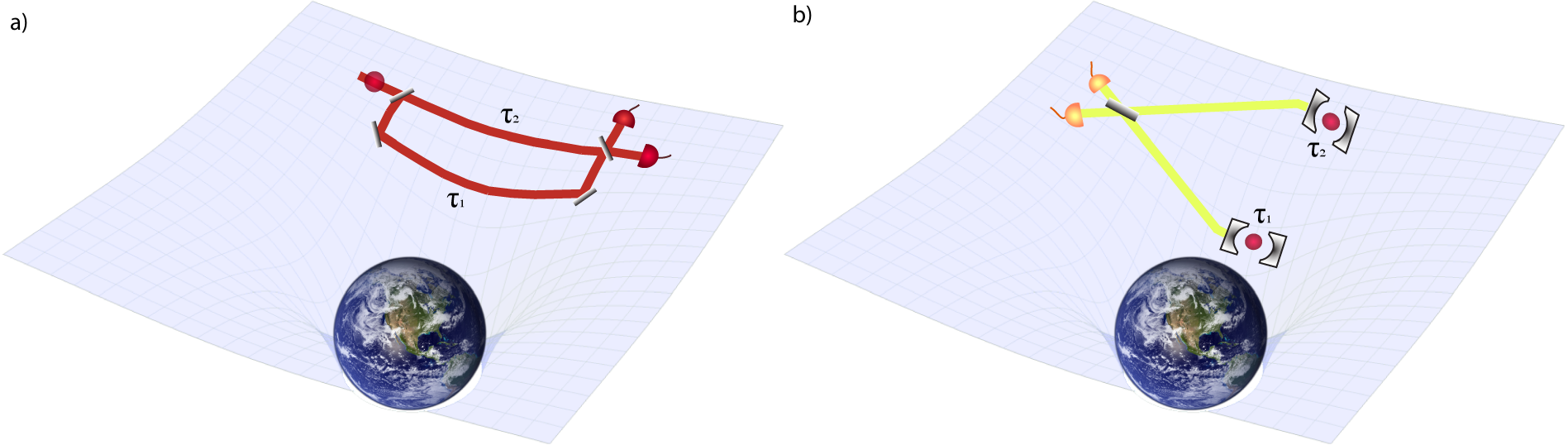}
\caption{Comparison of proper time interferometry with (a)  matter-waves and  (b) entangled clocks. The interplay between quantum theory and general relativity can be experimentally studied with a single clock in superposition, if the paths experience different proper times induced by the gravitational field with $\tau_1 < \tau_2$. The resulting entanglement between clock states and spatial degrees of freedom results in a modulation of the interferometric visibility. Here we show how such proper time interference can be realized with entangled clocks combined with non-local measurements enabled by photonic mediated entanglement distribution}.
\label{fig:ST}
	\end{figure*}

We now briefly review the essence of proper time interferometry \cite{Zych2011, pikovski2017time}: 
If a single clock follows two world lines in superposition, the two amplitudes will evolve separately according to their own experienced proper times $\tau_1$ and $\tau_2$. This is captured by the joint state of the clock and its superposed center-of-mass: $\ket{\Psi} = \frac{1}{\sqrt{2}} \left( U_c(\tau_1) \ket{c}\ket{\gamma_1} + U_c(\tau_2) \ket{c}\ket{\gamma_2}   \right)$, where $U_c(\tau_i)=e^{-i H_c \tau(\gamma_i)/\hbar}$ acts only on the clock states $\ket{c}$ and
 $\gamma_i$ are the two (fixed) trajectories of the center-of-mass of the system with corresponding proper times $\tau_i=\tau(\gamma_i)$ (Fig. \ref{fig:ST}). The path and the clock states become entangled due to the different proper-time evolutions. Creating and observing such situations in an actual experiment would demonstrate the interplay of gravitationally induced time dilation and the quantum superposition principle for proper-time evolutions. A signature of this interplay is the coherence: if the two paths are recombined and interference is observed, the recorded proper time in the clock states will provide which-way information, degrading the interferometric visibility. Tracing over the clock degrees of freedom, the observable is the visibility $V$ of interference, which is a direct measure of the proper time superposition:
 \begin{equation} \label{eq:Uinterference}
 V = |\langle U_c(\tau_1)U_c^{\dagger}(\tau_2) \rangle| = |\langle c(\tau_1)| c(\tau_2) \rangle| \, .
 \end{equation}
 Here the mean is taken with respect to the initial clock state $\ket{c}$.
As a concrete example, consider that the clock is initially localized and in the internal two-level superposition $\ket{c}=(\ket{a} + \ket{b})/\sqrt{2}$. We assume that the internal states have a relative energy difference of $\Delta E=hf$, where $f$ is the transition frequency between the two energy states and $h$ is Planck's constant. An example of such systems are atomic or ion clocks based on alkaline atoms where the two energy states correspond to two different electronic configurations~\cite{Aeppli2024,Zhang2022}.

After evolving in superposition along the two world-lines, upon recombination of the two amplitudes, it will 
yield the interferometric visibility for the spatial coherence given by
\begin{equation}\label{eq:visibilityMW}
V =|\langle U_c(\tau_1)U_c^{\dagger}(\tau_2) \rangle| = \left| \textrm{cos} \left( \frac{\Delta E (\tau_1-\tau_2)}{2 \hbar}\right) \right| \, .
\end{equation}
An observation of this modulation of coherence is thus a signature of the quantum nature of the proper-time evolution, which causes gravitationally-induced entanglement between clocks and paths that would not arise in Newtonian and classical physics. Observing this phenomenon thus amounts to a test of quantum dynamics with post-Newtonian corrections, which has so far no empirical evidence from experiments. 

 The outlined protocol has two major challenges. The first is practical: An experimental implementation is currently out of reach, since it requires a large coherent separation of matter-waves with additional internal clock states with large $\Delta E$. In the gravitational field on Earth, a second-long coherent separation of about $10$~m is required, combined with internal clock evolution on each amplitude at optical frequencies. The largest coherent matter-wave separations so far (without additional internal clock evolutions) are on the order of $0.5$~m for Kasevich-Chu interferometers \cite{kovachy2015quantum}, and on micrometer scales in lattice interferometers \cite{xu2019probing}.
The second challenge is of conceptual nature: While the experiment would probe the interplay of gravitational time dilation and quantum superposition, it would remain short of testing \textit{quantum dynamics on curved space-time}. This is because the gravitational time dilation in the lowest, homogeneous limit is equivalent to the special relativistic time dilation in an accelerated frame \cite{pikovski2017time, anastopoulos2018equivalence}, per the equivalence principle. It is thus desirable to push the experimental capabilities beyond this limit, such that no mapping to a special relativistic scenario can be performed. This would demonstrate a genuine test of quantum physics on curved space-time, a regime of physics that has so far not been experimentally accessible at all. 

\section{General description of proper time interference}
Proper time interference can be observed by other means than matter-wave experiments. To describe this, we will adapt an abstract view of clock interference in order to arrive at a general description of the phenomenon. For our purposes, we use the term `clock' to refer to a system with internal energy levels that are prepared in superposition, such that there is an internal time evolution. More generally it could be any system with internal degrees of freedom that are not in a single energy eigenstate, even non-periodic. For a general discussion of what constitutes a clock system, we refer the reader to Ref.~\cite{Milburn2020}.

Let's consider a four-dimensional Hilbert space that corresponds to the presence or absence of arbitrary clock systems at two different locations. Our basis vectors will thus be $\{\ket{nc,nc},\ket{nc,c},\ket{c,nc},\ket{c,c}\}$, where e.g. $\ket{nc,c}$ denotes the state with no clock system at position 1 and a clock at position 2. We now define a POVM (positive operator valued measure) consisting of 4 projectors:
\begin{equation} \label{eq:POVM1}
\mathbf{\Pi}=\{\Pi_{+},\Pi_{-},\Pi_{nn},\Pi_{cc}\},
\end{equation}
where $\Pi_{nn}=\ket{nc,nc}\bra{nc,nc}$, $\Pi_{cc}=\ket{c,c}\bra{c,c}$, and $\Pi_{\pm}=\ket{\pm}\bra{\pm}$ with $\ket{\pm}=\left(\ket{nc,c}\pm e^{-i\delta}\ket{c,nc}\right)/\sqrt{2}$.
This POVM describes a measurement that only reveals information about the presence/absence of clock systems and their spatial locations. What constitutes a clock system (or the absence of one) is still arbitrary: it can be the absence of the atom at the location altogether, or the inability to record time even if the atom is present. 
Time is recorded when the system is not in the eigenstate of the clock Hamiltonian $H_c$, thus at least two internal degrees of freedom are required such that its superposition forms a clock. Similarly, the absence of a clock system can be a state with only one internal degree of freedom such that no local time keeping is possible. In the matter-wave setup, we would have that $\ket{c}$ corresponds to the presence of an atom that can keep track of time, while $\ket{nc}$ would correspond to the absence of an atom i.e. vacuum.    

Now imagine that the clock system has an arbitrary number, $n$, of internal degrees of freedom such that it can be in any linear combination of $n$ orthonormal basis states $\{\ket{c_1},\ldots,\ket{c_n}\}$. However, the measurement described by the POVM in Eq.~(\ref{eq:POVM1}) still only reveals information about the presence/absence and spatial location of the clock systems - not their internal states. Including the extra degrees of freedom for the clock systems updates the POVM to
\begin{eqnarray} \label{eq:POVM2}
\Pi_{nn}\!\!\!&\to&\!\!\!\Pi_{nn} \nonumber \\
\Pi_{cc}\!\!\!&\to&\!\!\!\sum_{k,l=1}^n\ket{c_k}\bra{c_k}\otimes\ket{c_l}\bra{c_l} \nonumber \\
\Pi_{\pm}\!\!\!&\to&\!\!\!\frac{1}{2}\sum_{k=1}^n\Big(\ket{nc}\bra{nc}\otimes\ket{c_k}\bra{c_k}+\ket{c_k}\bra{c_k}\otimes\ket{nc}\bra{nc} \nonumber \\
&\!\!\!\pm&\!\!\!e^{i\delta}\ket{nc}\bra{c_k}\otimes\ket{c_k}\bra{nc}\pm e^{-i\delta}\ket{c_k}\bra{nc}\otimes\ket{nc}\bra{c_k}\Big). \nonumber \\
\end{eqnarray}
This general POVM captures the essence of proposed tests for quantum dynamics on curved space-time. For the matter-wave setup, we identified state $\ket{nc}_l$ ($\ket{c}_l$) as the state without (with) an atom at the $l$'th position. However, the clock can also be assigned to be a superposition between a subset of internal 
 ground states (hence referred to as stable states) of an atom with the absence of a clock corresponding to the atom being in a single orthogonal ground state, as we will see below. 
 The clocks may also be mixtures of excited states, which reproduces the results of time-dilation induced decoherence~\cite{Pikovski2015}. The key necessity for a successful implementation is that in such an experiment no classically well-defined proper time parameter can be assigned. This is achieved when measurements of $\Pi_{\pm}$ capture superpositions of different proper times.

We now consider a superposition state
\begin{equation} \label{eq:clock1}
\ket{\psi_0}=\frac{1}{\sqrt{2}}\left(\ket{nc}_1\ket{c(0)}_2+\ket{c(0)}_1\ket{nc}_2\right),
\end{equation}
at some initial reference time $t=0$. Here the indices $x=1,2$ refer to the two different locations of the two systems (in the following, we will omit writing these indices). As time ticks, the clock system will evolve differently depending on the gravitational field at the two locations resulting in a state  
\begin{equation}
\ket{\psi}=\frac{1}{\sqrt{2}}\left(\ket{nc}\ket{c(\tau_2)}+e^{i\Delta\Phi}\ket{c(\tau_1)}\ket{nc}\right), 
\end{equation}
where $\ket{c(\tau_x)}$ is the clock state evolved under the proper time at position $x$ and $\Delta\Phi=\Phi_1-\Phi_2$ with $\Phi_x=E_{nc}\tau_x/\hbar$ where $E_{nc}$ is the energy of the "no-clock" state.

We now perform the measurement corresponding to the POVM in Eq.~(\ref{eq:POVM2}). It is easy to verify that $\bra{\psi}\Pi_{cc}\ket{\psi}=\bra{\psi}\Pi_{nn}\ket{\psi}=0$, while
\begin{eqnarray} \label{eq:clock2}
\bra{\psi}\Pi_{\pm}\ket{\psi}&=&\frac{1}{2}\pm\frac{1}{4}\big(e^{i(\delta+\Delta\Phi)}\sum_k\bra{c(\tau_2)}\ket{c_k}\bra{c_k}\ket{c(\tau_1)} \nonumber \\
&+&e^{-i(\delta+\Delta\Phi)}\sum_k\bra{c(\tau_1)}\ket{c_k}\bra{c_k}\ket{c(\tau_2)}\big) \nonumber \\
&=&\frac{1}{2}\pm\frac{1}{2}|\bra{c(\tau_1)}\ket{c(\tau_2)}|\cos(\delta+\Delta\Phi+\lambda),
\end{eqnarray}
where $\bra{c(\tau_1)}\ket{c(\tau_2)}=|\bra{c(\tau_1)}\ket{c(\tau_2)}|e^{-i\lambda}$. We see the expected loss in visibility due to the different proper times at the two locations, measuring the same signature as in eq. \eqref{eq:Uinterference}.

We do not have to be restricted to the challenging spatial superposition state of a single atom as in matter wave interferometry. Instead, we can consider entangled states of two local quantum systems with at least three internal energy states. Imagine that we have two atomic quantum systems with internal states $\{\ket{g},\ket{a},\ket{b}\}$ with energies $E_g,E_{a}$, and $E_{b}$, respectively. We will now assign $\ket{nc}=\ket{g}$ and $\ket{c}=\alpha\ket{a}+\beta\ket{b}$ for arbitrary coefficients $\alpha$ and $\beta$. In other words, an atom at location $x$ in state $\ket{g}$ corresponds to the absence of a clock at position $x$ while an atom at location $x$ in any superposition of the states $\ket{a},\ket{b}$ corresponds to the presence of a clock at location $x$. The derivation from Eq.~(\ref{eq:clock1})-(\ref{eq:clock2}), will be exactly the same leading to the same oscillation of the visibility due to the difference in proper time.    

Importantly, both the preparation of the initial clock superposition (Eq.~(\ref{eq:clock1})) and the POVM measurement (Eq.~(\ref{eq:POVM2})) can be realized by means of entanglement distribution between the two locations and local detection. Auxiliary entangled pairs can be used to implement the non-local POVM and quantum networking techniques such as entanglement purification~\cite{Victoria2023} or even quantum repeaters~\cite{Azuma2023} can be employed to ensure faithful entanglement distribution in the presence of transmission loss and imperfections. Thus, tests of quantum dynamics on curved space-time can now take place on much larger scales utilizing quantum networks~\cite{pompili2021,Daiss2021,Knaut2024}.

\section{Entangled clock implementation}

\begin{figure} [t]
\centering
\includegraphics[width=0.49\textwidth]{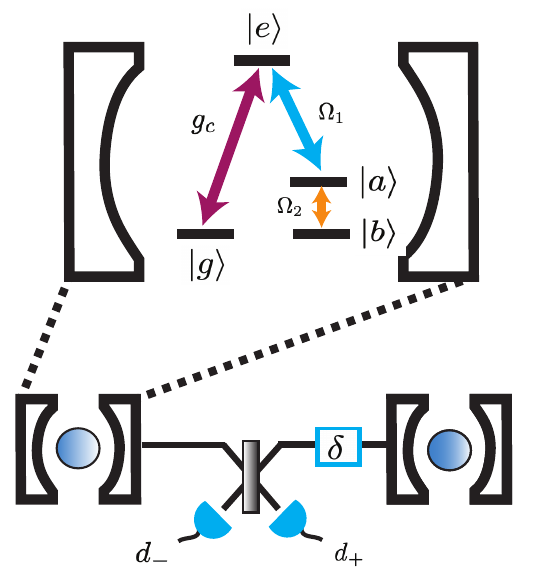}
\caption{Implementation of an entanglement-assisted test of general relativistic proper time in a quantum network. The transitions $\ket{a}\leftrightarrow\ket{e}$ and $\ket{a}\leftrightarrow\ket{b}$ can be driven by classical control fields denoted by $\Omega_1$ and $\Omega_2$, respectively. Emission from the $\ket{e}\to\ket{g}$ transition is collected with coupling $g_c$ to the collected mode depicted as an optical cavity mode. Note, however, that light could also be collected in free space. The collected light from the two emitters is interfered on a balanced beam splitter and a variable relative phase $\delta$ may be applied.}
\label{fig:setup}
	\end{figure}

As a specific example of proper time interference in a quantum network, we will consider the setup depicted in Fig.~\ref{fig:setup}. We focus on atomic or atom-like systems with three stable states labeled $\ket{g},\ket{a},\ket{b}$ and one excited state $\ket{e}$. The $\ket{a}\leftrightarrow\ket{e}$ transition can be driven by a classical pulse ($\Omega_1$), while single photons emitted on the $\ket{e}\leftrightarrow\ket{g}$ transitions are collected. In Fig.~\ref{fig:setup}, we have depicted the systems to be coupled to an optical resonator at resonance with the $\ket{e}\leftrightarrow\ket{g}$ transition for efficient light collection, however, this is not a requirement. Finally, we assume that arbitrary rotations in the $\ket{a},\ket{b}$ subspace can be achieved through another classical drive ($\Omega_2$). These two states will form our clock state $\ket{c} = \frac{1}{\sqrt{2}} \left(\ket{a} + \ket{b} \right)$.

\subsection{State preparation}
The first step of the protocol is to generate heralded entanglement between the two atomic systems by means of single photon emission and detection. For this purpose, we consider a single photon scheme~\cite{Zoller1999,Browne2003}, which is suitable for the considered level scheme.  

The atomic systems are initially prepared in state $\ket{a}$. By weakly driving the $\ket{a}\to\ket{e}$ transition on both atoms with phase-locked classical laser pulses, the systems can make the transition $\ket{a}\to\ket{g}\ket{1}_{ph}$ with a certain probability ($\epsilon$) that depends on the strength of the classical drive~\cite{Cirac1997,Boozer2007, Knall2022}. This will result in the state
\begin{eqnarray}
\ket{\psi'_0}&=&\sqrt{(1-\epsilon_1)(1-\epsilon_2)}\ket{a,a}\ket{0,0}_{ph} \nonumber \\
&+&\sqrt{\epsilon_1\epsilon_2}\ket{g,g}\ket{1,1}_{ph}+\sqrt{\epsilon_2(1-\epsilon_1)}\ket{g,a}\ket{1,0}_{ph} \nonumber \\
&+&\sqrt{\epsilon_1(1-\epsilon_2)}\ket{a,g}\ket{0,1}_{ph},
\end{eqnarray}
where $\ket{g,a}\ket{1,0}_{ph}$ denotes the first (second) atomic system in state $\ket{g}$ ($\ket{a}$) and a photon emitted from the first system.   

The emitted photons from both atomic systems are collected and sent to a balanced beam splitter. The excitation probabilities $\epsilon_1, \epsilon_2$ are tuned such that the probabilities of a photon from system 1 or 2 to be detected at the beam splitter is equal. This ensures that the beam splitter erases the "which-path" information of the photon such that the detection of a single photon after the beam splitter will ideally project the atomic systems into the entangled state $\ket{\psi}=\frac{1}{\sqrt{2}}\left(\ket{g,a}+\ket{a,g}\right)$ up to a single qubit phase correction depending on which detector records the photon.  Due to loss, there will be a trade off between rate and fidelity since a single photon detection can come from the state $|a,a\rangle$ if one of the photons are lost in transmission or non-number resolving detectors are used. This is treated in detail in App.~\ref{app:A} but for simplicity, we consider the evolution of the ideal state in the following.

After a successful entanglement generation attempt, the non-local clock is initialized. This is done by performing a $\pi/2$-rotation between the atomic levels $\ket{a}\leftrightarrow\ket{b}$ such that $\ket{a}\to\frac{1}{\sqrt{2}}\left(\ket{a}+\ket{b}\right)$ resulting in the state
\begin{equation}\label{eq:state1}
\ket{\psi'_0}\to|\psi_{0}\rangle=\frac{1}{2}\left(\ket{g}\left(\ket{a}+\ket{b}\right)+e^{i\theta_0}\left(\ket{a}+ e^{-i \varphi}\ket{b}\right)\ket{g}\right).
\end{equation}
 Here we also introduced an arbitrary relative phase shift $\varphi$ between the two atomic clocks, which can for example be implemented by initiating the two clocks at different times. We now have a non-local clock state similar to the state in Eq.~(\ref{eq:clock1}). The phase, $\theta_0=(E_g-E_{a})(T_1-T_2)/\hbar$ is the phase accumulated due to the difference in proper time ($T_1-T_2$) during the entanglement generation attempt. This phase is often negligible in realistic settings where each entanglement generation attempt is fast or the two states $\ket{g}$ and $\ket{a}$ are degenerate in energy. Nonetheless, we keep it for completeness and will later show that it will not affect the visibility measurement probing the interference of different proper time evolution.

After the start of the non-local clock, we let the state evolve for a time long enough that the the difference in proper time becomes non-negligible for the local clock super positions ($\ket{a}+\ket{b}$). This time will depend on the energy difference between the two internal clock states, i.e.  $\Delta E=E_{b}-E_{a}$.
 After the period of free evolution, the state will have evolved into
\begin{eqnarray} \label{eq:freeevol}
|\psi\rangle&=&\frac{1}{2}\big(\ket{g}\left(\ket{a}+e^{-i\theta_2}\ket{b}\right)\nonumber \\
&&+e^{i(\theta+\theta_0)}\left(\ket{a}+e^{-i\theta_1}\ket{b}\right)\ket{g}\big) \\
&=&\frac{1}{\sqrt{2}}\left(\ket{g}\ket{c(\tau_2)}+e^{i(\theta+\theta_0)}\ket{c(\tau_1)}\ket{g}\right) )\nonumber \, ,
\end{eqnarray}
where the phases are $\theta_1=\Delta E \tau_1/\hbar + \varphi$, $\theta_2=\Delta E \tau_2/\hbar$ and $\theta=(E_a-E_{g})(\tau_2-\tau_1)/\hbar$, and where we defined the clock states $|c(\tau_x)\rangle=(\ket{a}+e^{-i\theta_x}\ket{b})/\sqrt{2}$ $(x\in\{1,2\})$. 
Here $\tau_1$ and $\tau_2$ denote the proper times at the two locations.

\subsection{Measurement of the POVM}
In principle, auxiliary entangled pairs could be used to implement the non-local POVM in Eq.~(\ref{eq:POVM2}). One can naively think of this as teleporting the quantum state of the clock system at one location to the other using the auxiliary entanglement and then perform a local measurement. However, the POVM can also be realized by means of direct photonic interference if no auxiliary entanglement is available. A similar scenario using photon interference was considered in Ref. \cite{barzel2024entanglement} for tests of photonic redshift superpositions. Here we will use photonic interference as an example to show how the POVM realizing proper time interferometry is implemented.
After the free evolution to the state \eqref{eq:freeevol}, another $\pi/2$-pulse is applied to the levels $\ket{a}\leftrightarrow\ket{b}$. Imagine, e.g., that a reference signal is distributed between the two locations to ensure that the pulses are applied simultaneously. This transforms the state in Eq.~(\ref{eq:freeevol}) to
\begin{eqnarray}\label{eq:state2}
&&\frac{1}{2\sqrt{2}}\Big(\!\left(1\!+\!e^{-i\theta_2}\right)\ket{g,a}\!+\!\left(1\!-\!e^{-i\theta_2}\right)\ket{g,b}+\nonumber \\
&&e^{i(\theta+\theta_0)}\Big(\!\left(1\!+\!e^{-i\theta_1}\right)\ket{a,g}\!+\!\left(1\!-\!e^{-i\theta_1}\right)\ket{b,g}\!\Big)\!\Big) \qquad
\end{eqnarray}
  Ideally, a full state transfer from $\ket{a}\to\ket{g}\ket{1}_{ph}$ using a classical drive ($\Omega_1$) is now performed with each atomic system. 
This transforms the state in Eq.~(\ref{eq:state2}) to
\begin{eqnarray} \label{eq:state3}
&&\frac{1}{2\sqrt{2}}\Big(\ket{g,g}\Big(\!\left(1\!+\!e^{-i\theta_2}\right)\ket{0,1}_{E}\!\nonumber \\
&&+e^{i(\theta+\theta_0)}\!\left(1\!+\!e^{-i\theta_1}\right)\ket{1,0}_{E}\Big)\nonumber \\
&&+\Big(\!\left(1\!-\!e^{-i\theta_2}\right)\ket{g,b}\!\nonumber \\
&&+e^{i(\theta+\theta_0)}\!\left(1\!-\!e^{-i\theta_1}\right)\ket{b,g}\Big)\ket{0,0}_{E}\!\Big).
\end{eqnarray}
Here we we have used the notation $\ket{1,0}_E$ ($\ket{0,1}_E$) to denote the state with an (early) photon being emitted from system 1 (2). The populations in the $\ket{b}$ states are then transferred to the $\ket{a}$ states by a $\pi$-pulse between the levels and another state transfer, $\ket{a}\to\ket{g}\ket{1}_{ph}$ is performed. This transforms the state to
\begin{eqnarray} \label{eq:state4}
&&\frac{1}{2\sqrt{2}}\ket{g,g}\Big(\!\left(1\!+\!e^{-i\theta_2}\right)\ket{0,1}_E\ket{0,0}_L\!\nonumber \\&&+e^{i(\theta+\theta_0)}\!\left(1\!+\!e^{-i\theta_1}\right)\ket{1,0}_E\ket{0,0}_L\nonumber \\
&&+\!\left(1\!-\!e^{-i\theta_2}\right)\ket{0,0}_E\ket{0,1}_L\!\nonumber \\&&+e^{i(\theta+\theta_0)}\!\left(1\!-\!e^{-i\theta_1}\right)\ket{0,0}_E\ket{1,0}_L\!\Big),
\end{eqnarray}
where subscript $E$ ($L$) denotes an early (late) photon and we have assumed that the phase difference acquired during the time between the two photon emissions is negligible.

The early and late photons are being transmitted to the balanced beam splitter for detection with a controllable phase shift $\delta$ on one path. Letting the detector $d_{+}$ ($d_{-}$) detect the even (odd) superposition of the input modes the detection probabilities corresponding to Eq.~(\ref{eq:state4}) are
\begin{eqnarray}
P^{(E)}_{\pm}&=&\frac{1}{16}\abs{1+e^{-i\theta_2}\pm e^{i(\theta+\theta_0+\delta)}\left(1+e^{-i\theta_1}\right)}^2, \\
P^{(L)}_{\pm}&=&\frac{1}{16}\abs{1-e^{-i\theta_2}\pm e^{i(\theta+\theta_0+\delta)}\left(1-e^{-i\theta_1}\right)}^2,
\end{eqnarray}
where $P^{(E)}_{\pm}$ ($P^{L}_{\pm}$) is the probability of detecting the early (late) photon at detector $d_{\pm}$. 
The information of the arrival time of the photons that this measurement acquires does not correspond to the POVM in Eq.~(\ref{eq:POVM2}) since we obtain information about the internal states of the clock system. To correctly implement the POVM, we need to also interfere the early and late photons after the beam splitter, which can be achieved through additional interferometers and delay lines~\cite{Knaut2024}. However, since we only care about $\langle\Pi_{\pm}\rangle$ and not the post-measurement state, we do not need additional interferometers and can simply look at the total probability $P_{\pm}$ of detecting a photon at detector $d_{\pm}$, i.e. the sum of $P^{(E)}$ and $P^{(L)}$ since  
\begin{eqnarray}\label{eq:NLmeasure}
\langle\Pi_{\pm}\rangle&=&P_{\pm}^{(E)}+P_{\pm}^{(L)} \nonumber \\
&=&\frac{1}{4}\big(2\pm\cos(\delta+\theta+\theta_0) \nonumber \\
&&\pm\cos(\delta+\theta+\theta_0+\theta_1-\theta_2)\big) \nonumber \\
&=&\frac{1}{2} \pm \frac{1}{2}|\langle c(\tau_1)|c(\tau_2)\rangle|\cos(\delta+\Delta\Phi+\lambda),
\end{eqnarray}
where we have $\Delta\Phi=\theta_0+\theta$, and $\bra{c(\tau_1)}\ket{c(\tau_2)}=|\bra{c(\tau_1)}\ket{c(\tau_2)}|e^{-i\lambda}$.  
The probabilities $P_{\pm}$ will oscillate as the phase, $\delta$ is varied. The visibility of these oscillations will be
\begin{eqnarray}\label{eq:visibilityEnt}
\nu&=&\frac{\text{Max}_{\delta}(P_{\pm})-\text{Min}_{\delta}(P_{\pm})}{\text{Max}_{\delta}(P_{\pm})+\text{Min}_{\delta}(P_{\pm})} \\
&=&\frac{1}{2}(1+\cos(\theta_1-\theta_2)) \\
&= &  \cos^2 \left( \frac{\theta_1-\theta_2}{2} \right) \\
&= & \cos^2 \left( \frac{\Delta E (\tau_1-\tau_2) + \varphi}{2} \right) . \label{eq:vis}
\end{eqnarray}
The interference visibility is thus modulated due to the accumulated proper time differences, just as for matter-waves. For $\varphi=0$ the result is the same as in eq. ~\eqref{eq:visibilityMW}. The use of two entangled clocks now also enables this controllable phase shift which can tune the visibility, such as by $\pi$ for better scaling. Overall, we see in this concrete protocol that single photon mediated entanglement and non-local detection of two quantum emitters enables a test of quantum mechanics on curved space-time.   

\subsection{Imperfections and errors}
So far, we have neglected the effect of errors in the protocol. There will, however, be a number of imperfections for realistic experimental implementations. The most dominant will be transmission loss, finite photon collection and detection efficiencies, limited spin-coherence times, and interferometric instability in the setup.
The former three will affect the performance of both the entanglement generation and the readout step of the protocol. We note, however, that because both steps rely on photon detection, these errors will predominantly limit the success rate of the protocol since many errors can be heralded. This will not directly influence the quality of the measurements as long as the data-collection rate will be high enough to ensure a sufficient signal-to-noise ratio. For a more detailed discussion we refer to Appendix~\ref{app:A}.

\begin{figure} [t]
\centering
\subfloat{\includegraphics[width=0.49\textwidth]{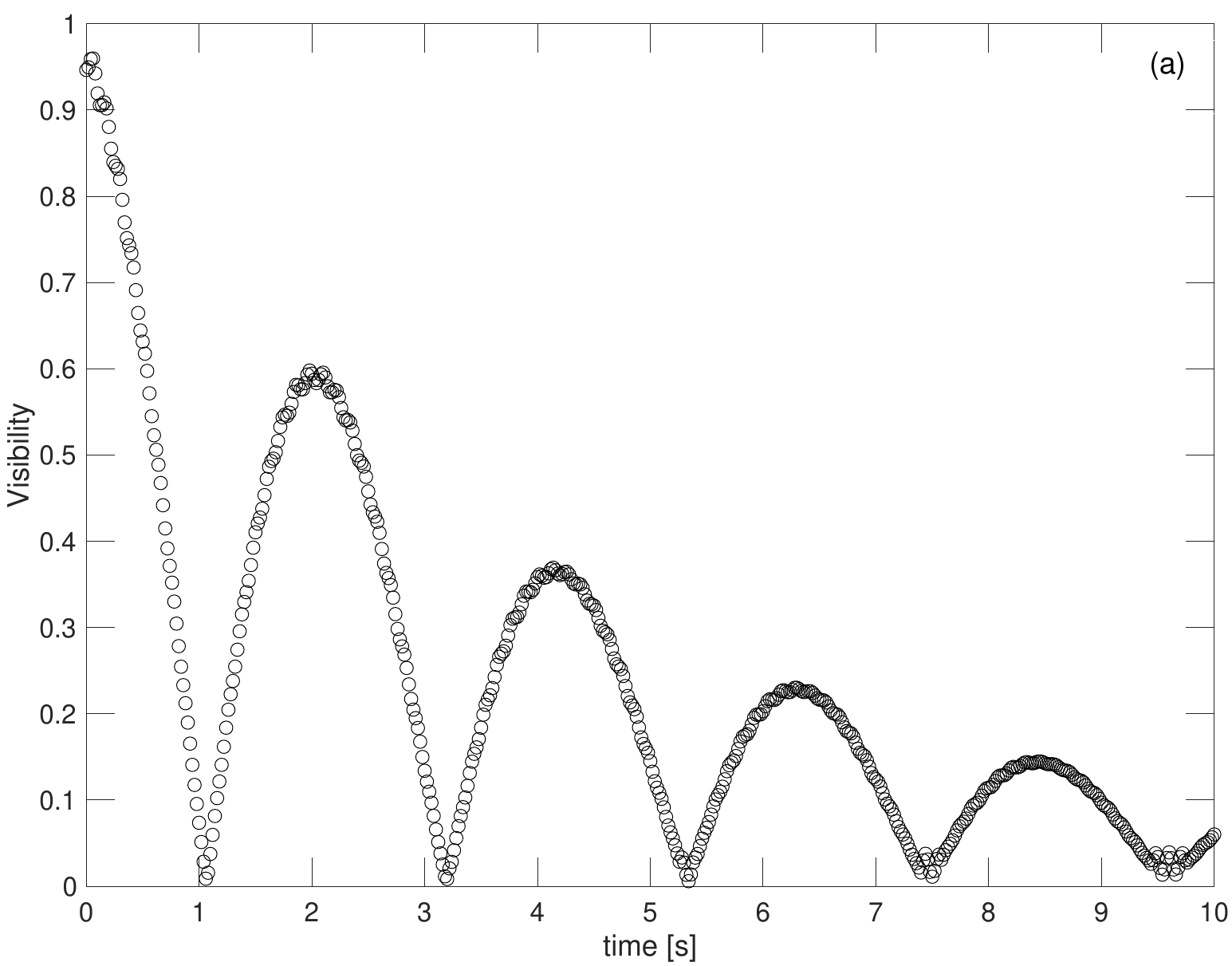}} \\
\subfloat{\includegraphics[width=0.49\textwidth]{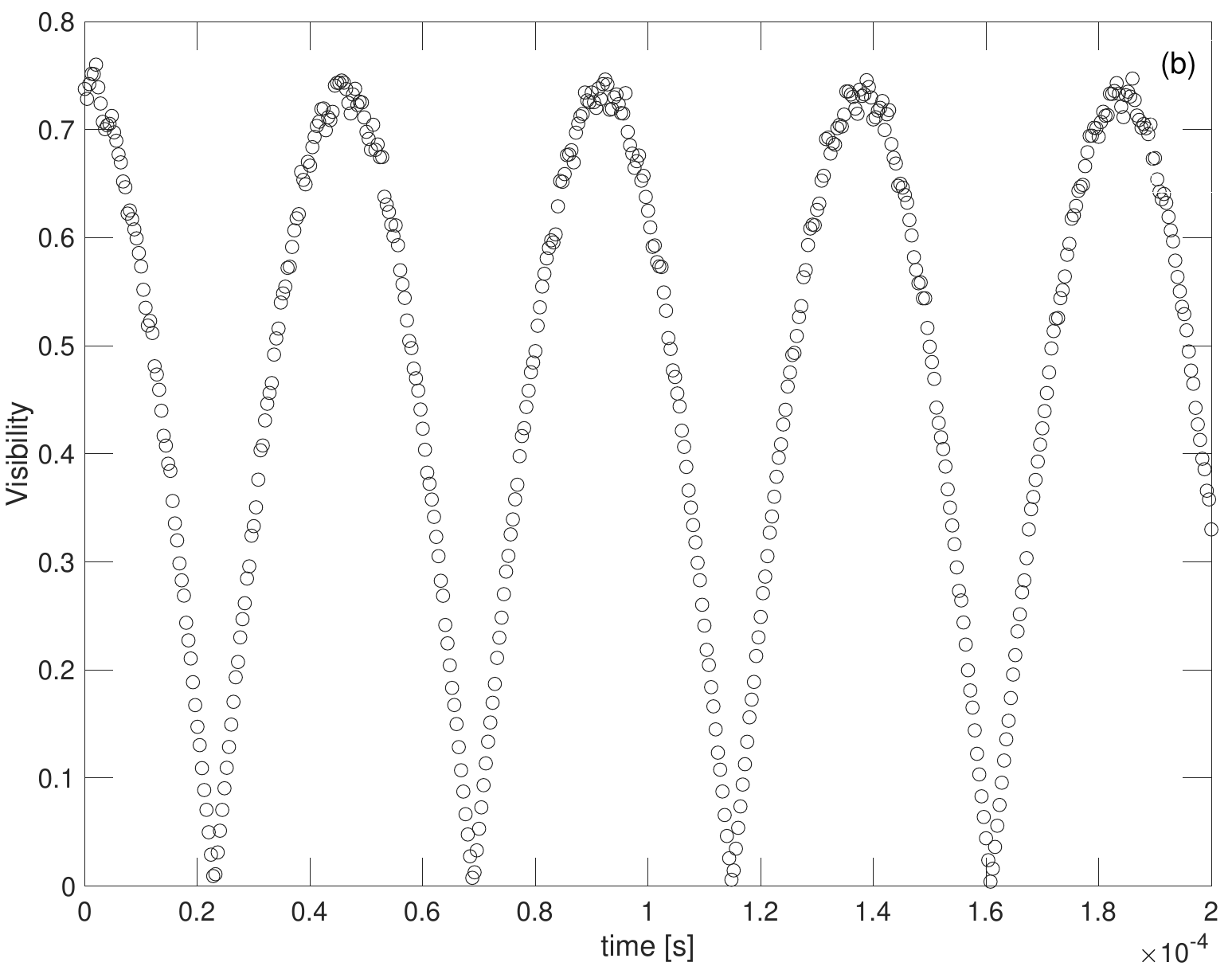}}
\caption{The visibility ($\nu$) as a function of the free evolution time for an implementation of the test with Strontium atoms. The simulations were performed with parameters detailed in App.~\ref{app:sim}. For (a) the height separation is 10 m and the spin coherence time is assumed to be 10s. Ih (b), we have simulated  an Earth-satellite setting where the height separation is 500km and the spin decoherence time is 0.1s.}
\label{fig:figure2}
\end{figure}

Errors from interferometric instability and limited spin coherence time will not be heralded and will directly degrade the quality of the measurement. The former will result in an overall reduction of the visibility while the latter will result in a decay of the visibility over time. This decay should be slower than the accumulation of relative phase from difference in proper time in order to measure a signal. For this reason, it is desirable to have a large energy separation between the clock states $\ket{a}$ and $\ket{b}$. The long coherence times (seconds) of optical transitions in Strontium and Ytterbium atoms and possibility to couple to nanophotonic cavities for efficient photon collection~\cite{Robinson2024,Covey2019} makes such atomic species particularly suited for a test of gravitational time dilation. Consequently, the visibility oscillations due to gravitational time dilation could be observed on a times scale of seconds with only a 10m height difference between the two stations and a 10 second long spin coherence time as shown in Fig.~\ref{fig:figure2}(a) where the visibility as a function of time is simulated for Strontium atoms.   

The photon mediated entanglement setup is also compatible with Earth-satellite configurations, which allows to probe richer space-time geometries beyond surface effects. Thus, while quantum payloads have been envisioned for quantum communication~\cite{Liao2018}, it may also enable new fundamental tests of gravity and quantum mechanics. Focusing on a \emph{low-Earth-orbit} satellite, Fig.~\ref{fig:figure2}(b) shows the simulated visibility signal as a function of time assuming a height difference of 500 km and a more modest decoherence time of 0.1 second of Strontium atoms.   

While our analysis focused on a specific implementation utilizing single photon mediated entanglement generation and interferometric readout in Strontium atoms, other schemes can also be suitable depending on the specific hardware.  The main requirement is that they allow for distributed entanglement and measurement. Examples of other possible systems include Ytterbium atoms \cite{safronova2018two}, entangled BECs \cite{cassens2024entanglement} or color vacancy centers in diamonds \cite{ruf2021quantum}. The use of separate but entangled clocks opens a range of opportunities not only in terms of hardware, but also for state preparation. As mentioned above, one can tune the relative phase $\varphi$ between the clocks for a more favorable scaling in the visibility \eqref{eq:vis}. Another possibility is to use GHZ-states of $N$ clocks~\cite{Covey2024}: this effectively changes the relative  frequency accumulation by $N\Delta E$, which can alleviate other experimental paremters, although the generation of such entangled states remains challenging. Nevertheless, it shows how quantum networks have unique advantages and variability for tests of quantum theory on curved space-time.
 
\section{Measurement-based implementation}
In our discussion above we highlighted the use of the POVM $\{\Pi_{+},\Pi_{-},\Pi_{nn},\Pi_{cc}\}$ in order to observe superpositions of different proper time evolutions, and how it can be implemented with entangled clocks. The non-local measurement of $\langle\Pi_{\pm}\rangle$ contains the proper time interference term $U_c(\tau_1)U_c^{\dagger}(\tau_2)$ as in eq.~\eqref{eq:Uinterference}. In fact, this highlights that the non-local measurement is the key, rather than the initial state preparation. Within regular quantum theory, the same result~\eqref{eq:NLmeasure} can also be obtained by \textit{post-selecting} the evolved state through measurement. This opens the possibility to perform the protocol even with two initially non-entangled atoms, which can be projected onto an entangled subspace by the non-local interference measurement. We demonstrate this by starting the two clocks in superposition of \textit{three} levels, and an initial product state $\frac{1}{3}(\ket{g}+\ket{a}+\ket{b})\otimes(\ket{g}+\ket{a}+\ket{b})$. The clocks will evolve according to their respective free evolutions $U_c(\tau_1)$ and $U_c(\tau_2)$, respectively. In particular, the state becomes 
\begin{equation}
    \begin{split}
    \ket{\psi} & =\frac{1}{3} e^{-i \frac{E_g}{\hbar}(\tau_1+\tau_2)} \ket{g}\ket{g} + \frac{2}{3} e^{-i \frac{E_a}{\hbar}(\tau_1+\tau_2)}\ket{c(\tau_1)}\ket{c(\tau_2)} \\
    & + \frac{\sqrt{2}}{3} e^{-\frac{i}{\hbar} (E_g\tau_1+ E_a\tau_2)} \left( \ket{g}\ket{c(\tau_2)} + e^{i \theta} \ket{c(\tau_1)}\ket{g}  \right) \, .
    \end{split}
\end{equation}
This of course is still a product state but the measurement of $\Pi_{\pm}$ picks up the entangled contributions in the last term, post-selecting on exactly the state~\eqref{eq:freeevol}. 
Thus the same interference of the two evolutions can be obtained when post-selecting on the outcomes of the $\Pi_{+},\Pi_{-}$ measurements, resulting in 
\begin{equation} \label{eq:Vps}
    \nu_{ps} = |\bra{c(\tau_1)}\ket{c(\tau_2)}_{ps}| \, .
\end{equation} 
$\nu_{ps}$ is the resulting visibility due to proper time interference, which is the same as $\nu$ in eq. \eqref{eq:visibilityEnt} for an initial entangled state. While the product state approach might be experimentally easier since it does not require the initial preparation of an entangled state, it still requires a non-local measurement, which in general requires entanglement distribution. 

Conceptually, however, this protocol tests a subtly different aspect of proper time evolution in quantum theory.  While it yields the same outcome as the initially entangled state, $\nu_{ps} = \nu$, this is based on the application of regular quantum theory, and in particular the linearity and basis-independence. The measurement reveals that in normal quantum theory, even the product state can be decomposed in an entangled basis, and the non-local measurement picks up these contributions. It is necessary to discard measurements that correspond to the other branches of the state, and only post-select to the outcomes that would correspond to \eqref{eq:freeevol}, which here would occur with probability $p=4/9$. Taking the other measurements of the POVM into account one can confirm that the state is a product state. The result thus post-selects on the entangled amplitudes of the unitary, linear evolution which are necessary to reproduce the product state evolution in this particular basis. In contrast, an initial entangled state evolves dynamically under different proper times in coherent superposition as a whole, and the entanglement can be confirmed independently for the state throughout the protocol.   
Within regular quantum theory, there is no  difference between these two scenarios for the measurement we propose, and the outcome is the same: the result reveals $U_c(\tau_1)U_c^{\dagger}(\tau_2)$ deterministically for the entangled case, and with some finite probability in the case of post-selection for an initial product state. The agreement between the two is fundamentally linked to the fact that proper time is sourced by a classical background space-time, and is thus a fixed classical background, as opposed to a superposition of different proper time evolutions even at the same location (which would amount to a quantum gravity scenario of superpositions of space-times). However, for a \textit{test} of quantum dynamics on curved space-time, the two cases are conceptually different. One involves the superposed dynamics of the full non-classical state, while the other projects the classical state onto a non-local superposition and reveals these coherently superposed amplitudes of the state. Within an alternative framework, these two scenarios could lead to different results such that $\nu_{ps} \neq \nu$. In this case the expected quantum dynamics of superposed proper times does not proceed according to the usual formalism, for which there is yet to be empirical evidence. For example, a non-linear and/or non-unitary modification in the presence of space-time curvature could lead to different outcomes, since then evolutions of individual amplitudes in different bases might evolve differently. One can therefore envision a more comprehensive test consisting of both the approaches we described: initially entangling two clocks and evolving them with respect to different proper times, then measuring $\nu$ in eq.~(\ref{eq:visibilityEnt}) using the protocol outlined in the previous section; and comparing the result to evolving two independent clocks with their respective proper times, and then performing a non-local measurement to obtain $\nu_{ps}$ in eq.~(\ref{eq:Vps}). We quantify any possible discrepancy by the parameter  
\begin{equation}
\lambda = \nu_{ps} - \nu \, .
\end{equation}
Both approaches should yield the same outcome according to the expected quantum dynamics on curved space-time for which $\lambda=0$. But if linearity, unitarity, or even the Born rule no longer hold in curved space-time, such as in some models and considerations \cite{hawking1982unpredictability,sorkin1994quantum, ralph2014entanglement,berglund2022gravitizing,hubsch2024quantum}, then $\lambda \neq 0$ could occur. Confirming or constraining this parameter would thus be a new test of quantum principles on curved space-time, an unexplored frontier of quantum physics.

\section{Conclusion and Outlook}
In conclusion, we have proposed a test of the interplay between general relativity and quantum mechanics based on quantum networks with entangled clocks and non-local measurements. We showed that such tests can probe how different gravitationally induced proper times interfere, and how the resulting entanglement modulates the interferometric visibility of non-local measurements through emitted photons. Our scheme extends previously considered matter-wave scenarios to entangled systems, allowing for tests across different space-time curvatures. Optical atomic clocks such as Strontium and Ytterbium atoms offer a promising route for implementation with current state-of-the-art, due to their long coherence times and stable optical transitions. As we have demonstrated here, joint measurements in an entangled basis is sufficient to study the quantum mechanical interference of proper time. This opens the possibility to perform such tests also with initially unentangled atomic clocks and non-local measurements, and to probe the predicted equality with the entangled case as another test of the predictions of quantum theory on curved space-time. Our framework can also easily be extended to study interference of multiple points in the gravitational field by considering e.g. W-like entangled states~\cite{Pu2018} and generalized basis measurements and opens up a new avenue for using quantum networks to perform fundamental tests of nature that are impossible with classical sensors.

\subsection*{Acknowledgements}
We thank Jacob Covey, Marianna Safronova, James Thompson and Jun Ye for valuable discussions. This work was supported by the NWO Gravitation Program Quantum Software Consortium (Project QSC No. 024.003.037), the AWS Quantum Discovery Fund at the Harvard Quantum Initiative, the National Science Foundation under grant 2239498, NASA under grant 80NSSC25K7051 and by the Alfred P. Sloan Foundation under grant G-2023-21102. 


\appendix
\section{Simulation of protocol} \label{app:A}
Here we describe our simulation of the protocol for testing the effect of gravitational time-dilation on quantum systems. The protocol consists of three steps: Entanglement generation, free evolution, and readout.

\emph{Entanglement generation:} We assume that the the two stations contain identical quantum systems that are initially prepared in the ground state $\ket{a}$ with efficiency $\eta_i$. The initial state for each system is modeled as
\begin{equation}
\rho_0=\eta_i\ket{a}\bra{a}+\frac{(1-\eta_i)}{2}\left(\ket{b}\bra{b}+\ket{g}\bra{g}\right),
\end{equation}
where we have assumed that with probability $(1-\eta_i)$, the system ends up in an equal mixure of the two other stable states. A weak laser pulse drives the transition $\ket{a}\to\ket{e}$ resulting in a small probability for a photon to be emitted from the excited state such that the systems ends in the ground state $\ket{g}$. The excited state might, however, also decay by emitting an uncollected photon or by some non-radiative decay.  We model this process as a transformation $\ket{a}\to\sqrt{\epsilon}\left(\sqrt{1-p_c}\ket{\tilde{o},0}+\sqrt{p_c}\ket{g,1}\right)+\sqrt{1-\epsilon}\ket{a,0}$ described by a unitary $U_{c}$. Here $\ket{g,1}$ is where the atomic system is in spin-state $\ket{g}$ and a single photon has been emitted into the collected mode. The generic noise state $\ket{\tilde{o},0}$ is the state following a spontaneous emission where no collected photon is emitted. After tracing out the environment (containing e.g the non-collected photons or the non-radiative decay channels), the resulting density matrix is
\begin{equation}
\rho_1=\rho_c+\rho_{\tilde{o}},
\end{equation}
where $\rho_c=(\mathbb{I}-P_{\tilde{o}})U_c\rho_0U^{\dagger}_c(\mathbb{I}-P_{\tilde{o}})^{\dagger}$ and $\rho_{\tilde{o}}=P_{\tilde{o}}U_c\rho_0U^{\dagger}_cP_{\tilde{o}}^{\dagger}$ are the unnormalized density matrices resulting from the emission of a collected photon and non-collected photon, respectively. Here, $\mathbb{I}$ is the identity matrix and $P_{\tilde{o}}=\ket{\tilde{o},0}\bra{0,\tilde{o}}$. We then assume for simplicity that the noise state corresponds to an equal mixture of all stable states e.g. that $\ket{\tilde{o}}\bra{\tilde{o}}=\left(\ket{g}\bra{g}+\ket{a}\bra{a}+\ket{b}\bra{b}\right)/3$. 

The outcoupling efficiency and transmission loss to the central station is modeled as a fictitious beam splitter with transmission $\eta_o\eta_t$ that mixes the signal (photonic part of $\rho_1$) with vacuum. In order to model interferometric instability, we also apply a unitary transformation $U_{\text{ins}}=\mathbb{I}-(1-e^{i\xi_i})\ket{g,1}\bra{1,g}$ to $\rho_1$, where $\xi_i$ is a random phase. Let $\rho_2$ be the density matrix describing the system following this photon loss and phase accumulation. The input state to the symmetric beam splitter at the central station is then $\tilde{\rho}_0=\rho^{(1)}_2\otimes\rho^{(2)}_2$ where the superscript of $\rho_2^{(1)}$ ($\rho_2^{(2)}$) denotes the two different atomic systems. Note that in general $\rho_2^{(1)}\neq\rho_2^{(2)}$ since the different parameters such as excitation probability, propagation phase, etc. of the two atomic systems may be different.

After the beam splitter, the two output ports are measured with single photon detectors.  We assume that the photo detectors are not number resolving and that they have a detection efficiency of $\eta_d$. Furthermore, we neglect dark counts, allowing us to write the resulting (unnormalized) density matrix following a single click in one of the detectors as
\begin{eqnarray}
\rho_3&=&\eta_d(P_{10}U_{\text{BS}}\tilde{\rho}_0U_{\text{BS}}^{\dagger}P_{10}+U_{\pi}P_{01}U_{\text{BS}}\tilde{\rho}_0U_{\text{BS}}^{\dagger}P_{01}U_{\pi}^{\dagger})+ \nonumber \\
&&(\eta_d^2+2\eta_d(1-\eta_d))(P_{20}U_{\text{BS}}\tilde{\rho}_0U_{\text{BS}}^{\dagger}P_{20}+\nonumber \\
&&U_{\pi}P_{02}U_{\text{BS}}\tilde{\rho}_0U_{\text{BS}}^{\dagger}P_{02}U^{\dagger}_{\pi}).
\end{eqnarray}
Here $P_{10}$ ($P_{01}$) is the projection on to states containing one photon in output port 1 (2) and similarly two photons for $P_{20}$ ($P_{02}$). The central beam splitter transformation is described by a unitary $U_{\text{BS}}$, which makes the transformations $\ket{1,0}\to(\ket{1,0}+\ket{01})/\sqrt{2}$, $\ket{0,1}\to(\ket{1,0}-\ket{01})/\sqrt{2}$, and $\ket{1,1}\to(\ket{2,0}-\ket{0,2})/\sqrt{2}$ where $\ket{1,0}$ denotes a photon in port 1 and vacuum in the other port. Note that if a photon is measured in port 2, a phase correction described by $U_{\pi}$, which amounts to a $\pi$-phase shift of the $\ket{a}$ spin state of the second atomic system. This concludes the entanglement generation step of the protocol and the success probability of this step is $P_s=\Tr{\rho_3}$. For small imperfections $P_s\approx\eta_d\eta_o\eta_t\epsilon$, while the fidelity the target state $\ket{\psi}$ (see main text) is $F\sim1-2(1-\eta_d\eta_o\eta_t)\epsilon$~\cite{Borregaard2015} .

While the local operations are usually fast ($\mu$s timescale), the signalling time between the atomic systems might be long for large separations (e.g. Earth-satellite). We  therefore include limited coherence time of the spin systems by applying a depolarizing channel of the form 
\begin{equation}
\Lambda_t(\rho)=(1-p)\rho+\frac{p}{3}(\ket{g}\bra{g}+\ket{a}\bra{a}+\ket{b}\bra{b})
\end{equation}
where $p=\text{Exp}[-t/T_{d}]$ is the probability of the state being depolarized after time $t$ determined by an effective decoherence time $T_d$. We assume that the time of an entanglement generation attempt is step by the signalling time $T_c$ between the two atomic systems. The resulting density matrix is then $\rho_4=\Lambda_t(\rho_3)/P_s$

\emph{Free evolution:} We continue to model the free evolution part of the protocol. First a $\pi/2$-pulse described by a unitary $U^{(x)}_{\pi/2}$ on each atomic system at location $x$ implements the transformation $\ket{a}_x\to\sqrt{\Omega_x}\ket{a}_x+\sqrt{1-\Omega_x}e^{i\phi_{\pi,x}}\ket{b}_x$. Subsequently evolution during the period, $T$ of free evolution can be described by local unitaries
\begin{equation}
U^{(x)}_T=e^{-i\theta^{(x)}_g}\ket{g}_x\!\!\bra{g}+e^{-i\theta^{(x)}_{c1}}\ket{a}_x\!\!\bra{a}+e^{-i\theta^{(x)}_{c2}}\ket{b}_x\!\!\bra{b},
\end{equation}
acting on the spin systems. Here $\theta^{(x)}_g=\omega_gT(1+\phi_x/c^2)$ with $\phi_x$ being the gravitational potential at site $x$ and similarly for $\theta^{(x)}_{c1}$ and $\theta^{(x)}_{c2}$. Thus, the density matrix after the period of free evolution is
\begin{eqnarray}
\rho_{5}&=&(U^{(1)}_T\otimes U^{(2)}_T)(U^{(1)}_{\pi/2}\otimes U^{(2)}_{\pi/2})\rho_{4}\nonumber \\
&&(U^{(1)}_{\pi/2}\otimes U^{(2)}_{\pi/2})^{\dagger}(U^{(1)}_T\otimes U^{(2)}_T)^{\dagger}.
\end{eqnarray}
To model the spin decoherence during the period of free evolution, we also apply the same single atom depolarizing channel as in the entanglement generation step but for a time of $T$ in stead of $T_c$. At the end of the free evolution step, the total density matrix describing the state of the two spin systems is thus
$\rho_{6}=\Lambda_T(\rho_5)$.

\emph{Readout:} The readout is modelled in a very similar way as the entanglement generation step. The local rotations of the clock states $\{\ket{a},\ket{b}\}$ are modelled with unitaries similar to $U^{(x)}_{\pi/2}$. The photon emission is modeled as a local unitary $U^{(x)}_{p}$, which makes the transformation $\ket{a}_x\to\sqrt{1-p_c'}\ket{\tilde{o},0}_x+\sqrt{p_c'}\ket{g,1_E}_x$. Here $\ket{g,1_E}$ is the state where the spin is in the ground state $\ket{g}$ and an early (collected) photon has been emitted. With probability $1-p_c'$, no photon is collected and the spin is prepared in the generic noise state $\ket{\tilde{o},0}_x$. As in the entanglement generation step, we replace the population in this state with an equal mixture of all ground states. 

Subsequently a $\pi$-pulse makes the transformation $\ket{b}\to\ket{a}$ and a second state transfer described by the transformation $\ket{a}\to\sqrt{1-p_c'}\ket{\tilde{o}',0}_x+\sqrt{p_c'}\ket{g,1_L}_x$, where $\ket{1_L}$ is a late (collected) photon is performed. We again replace the generic noise state with an equal mixture of all ground states.

Outcoupling and transmission loss together with interferometric instability is modeled as in the entanglement generation step with fictitious beam splitters and a fluctuating phase ($\xi'_i$) on the photon states (assumed the same for the early and late photon). Note that a controllable phase $\delta$ is also imposed on one of the arms of the interferometer (see Fig.~\ref{fig:setup}). The central beam splitter as assumed perfect as in the entanglement generation step. To calculate the probability of obtaining a click in detector $d_+$ or $d_-$, we again neglect dark counts, assume a detection efficiency of $\eta_d$ and that the detectors are not number resolving. We do, however assume that they can distinguish between early/late clicks and we post-select on only one detector recording a click in either the early or late time interval.

\section{Simulation details} \label{app:sim}

We assume that both state $\ket{g}$ and $\ket{a}$ are encoded as two different hyperfine states of the $1\text{S}_0$ ground manifold of $^{87}$Sr atoms while the $\ket{b}$ is encoded in the $3\text{P}_0$ manifold~\cite{Robinson2024,Barnes2022}. Consequently, the clock transition is $698$ nn and the states $\ket{g}$ and $\ket{a}$ are approximately degenerate in energy. The simulations in Fig.~\ref{fig:figure2} are performed with parameters:

\begin{itemize}
\item $\eta_i=0.99$, $\epsilon=0.1$
\item $\eta_o=0.5$, $\eta_d=0.5$, 
\item $\Omega_a=0.5$, $\theta_{\pi,a}=0$
\end{itemize}

These parameters are compatible with current neutral atom experimental hardware. Initialization errors and single qubit rotations with fidelities of $>99\%$ have been demonstrated with several atomic species~\cite{Xia2015,Ma2022} as well as atom-atom entanglement with fidelities of $>98\%$ for separations of 21 m~\cite{Ritter2012}.  
The interferometric instability is modeled by picking random values of $\xi'_i$ according to a Gaussian distribution centered at a mean value. The format $\xi'_i=0\pm0.1$ means a mean of $0$ and standard deviation of $0.1$. For the simulations in Fig.~\ref{fig:figure2}(a), we have used $\xi'_i=0.0\pm0.1$, while for Fig.~\ref{fig:figure2}(b) we used $\xi'_i=0.0\pm 1.0$ as example parameters. The difference in phase variability is to reflect the different environment of a fiber based and a free space (satellite) link. For the former, recent metropolitan quantum networks have demonstrated phase-locking to the \% error level for entanglement generation~\cite{stolk2024} compatible with the assumed value in the simulation. For satellite-based implementations, phase stability is more demanding due to atmospheric effects such as turbulence. However, fringe tracking techniques and adaptive optics can be used to track the phase and minimize the phase fluctuations~\cite{wu2024single}. Additionally, entanglement schemes based on two-photon interference~\cite{Bernien2013} could be employed to have lower sensitivity to phase fluctuations than the single-photon scheme considered in this work. The transmission $\eta_t$ was assumed to be $0.9$ and $0.01$ for Figs.~\ref{fig:figure2}(a) and (b), respectively.       

%

\end{document}